\documentstyle[hx37_latex]{article}

\def\spose#1{\hbox to 0pt{#1\hss}}

\def\kms{\ifmmode {\rm\,km\,s^{-1}}\else ${\rm\,km\,s^{-1}}$\fi}

\def\kmsmpc{\ifmmode {\rm\,km\,s^{-1}\,Mpc^{-1}}\else ${\rm\,km\,s^{-1}\,Mpc^{-1}}$\fi}

\def\ergps{\ifmmode {\rm\,erg\,s^{-1}}\else ${\rm\,erg\,s^{-1}}$\fi}
\def\ergpscm2{\ifmmode {\rm\,erg\,s^{-1}\,cm^{-2}}\else
    ${\rm\,erg\,s^{-1}\,cm^{-2}}$\fi}

\def\deg{\ifmmode {^{\circ}}\else {$^\circ$}\fi}
\def\degr{\ifmmode {^{\circ}}\else {$^\circ$}\fi}
\def\degs{\ifmmode {^{\circ}}\else {$^\circ$}\fi}

\def\etal{{\it et al.~}}

\def\h3Mpc{h^{-3}{\rm Mpc}^3}

\def\arcsec{\ifmmode {^{\prime\prime}}\else $^{\prime\prime}$\fi}
\def\asec{\ifmmode {^{\prime\prime}}\else $^{\prime\prime}$\fi}
\def\arcmin{\ifmmode {^{\prime}}\else $^{\prime}$\fi}
\def\amin{\ifmmode {^{\prime}}\else $^{\prime}$\fi}

\def\secper{\ifmmode \rlap.{^{s}}\else $\rlap{.}{^{s}} $\fi}
\def\minper{\ifmmode \rlap.{^{m}}\else $\rlap{.}{^m} $\fi}
\def\secspt{\ifmmode \rlap.{^{\prime\prime}}\else
    $\rlap.{^{\prime\prime}}$\fi}
\def\arcsper{\ifmmode \rlap.{^{\prime\prime}}\else
    $\rlap.{^{\prime\prime}}$\fi}
\def\minspt{\ifmmode \rlap.{^{\prime}}\else
    $\rlap.{^{\prime}}$\fi}
\def\arcmper{\ifmmode \rlap.{^{\prime}}\else
    $\rlap.{^{\prime}}$\fi}
\def\spose#1{\hbox to 0pt{#1\hss}}
\def\simlt{\mathrel{\spose{\lower 3pt\hbox{$\mathchar"218$}}
     \raise 2.0pt\hbox{$\mathchar"13C$}}}
\def\simgt{\mathrel{\spose{\lower 3pt\hbox{$\mathchar"218$}}
     \raise 2.0pt\hbox{$\mathchar"13E$}}}

\def\refindent{\par\noindent\parskip=2pt\hangindent=3pc\hangafter=1 }

\def\ref#1;#2;#3;#4 {\refindent{#1,} {#2}, #3, #4}

\def\book#1;#2;#3 {\refindent{#1, }{in {\it{#2},} }{#3}}

\input{psfig}
\def\be{\begin{equation}}
\def\ee{\end{equation}}
\def\bea{\begin{eqnarray}}
\def\eea{\end{eqnarray}}

\begin{document}

\title{CLUSTERS OF GALAXIES AT $z>1$}

\author{ MARK DICKINSON }

\address{STScI, 3700 San Martin Dr., Baltimore MD 21218 USA}

\maketitle\abstracts{
Although field galaxy studies have begun to probe the universe at
$z > 1$, evidence for galaxy clusters at such redshifts
has been sparse.  New observations are accumulating rapidly, however,
providing new data on the early evolution of elliptical galaxies, 
the blue ``Butcher--Oemler'' population, and of large scale structure
at unprecedentedly large lookback times.  I briefly review some of
these new observations, discussing morphological and spectral 
characteristics of cluster galaxies at $z > 1$, x--ray evidence
for massive, collapsed clusters out to $z=1.8$, and tantalizing
indications from the literature for clusters at $2 < z < 5$.
}

\section{Introduction}

Until quite recently, a talk with the title given above would have
been either exceedingly short or entirely theoretical.  Although rich
clusters of galaxies have been popular and productive laboratories for 
studying galaxy evolution for many years (cf. the contributions of 
Dressler and others to this volume), published surveys for distant clusters 
largely run out of steam by $z = 1$.  The most distant clusters with 
measured redshifts in optical or x--ray surveys have $z \approx 0.9$.
While this might reflect an evolution in the cluster space density
(perhaps there simply are no rich clusters at $z > 1$?), it is more
probably an effect of observational selection.  At $z > 1$, 
strong k--corrections, particularly for the red, early--type galaxies 
which dominate nearby clusters, may greatly reduce the visibility of 
distant cluster galaxies when observed at optical wavelengths, 
and the contrast of even a rich cluster against the tremendously numerous 
population of faint field galaxies
may be diminished to the point of near invisibility.  

From my perspective, the search for clusters beyond the limits of present 
surveys is primarily motivated by three interests:  (1) the hunt for the 
formation epoch of early--type cluster galaxies, (2) the early evolution of 
the ``Butcher--Oemler'' blue cluster population, and (3) evidence for
massive, collapsed clusters at high redshift as a constraint on cosmology
and theories of large scale structure formation.  I will touch upon all
of these points briefly below, and provide a very brief summary of the
most enticing evidence from the literature for galaxy clusters at $z > 2$.  
Because of space constraints, I primarily limit my 
discussion to clusters identified as collections of {\it galaxies;}
further evidence for cluster{\it ing} at high redshift (e.g. from QSO 
absorption lines, quasar angular correlations, etc.) will unfortunately
be neglected here.

\section{Clusters at $z < 2$}

\begin{figure}
\centerline{\psfig{figure=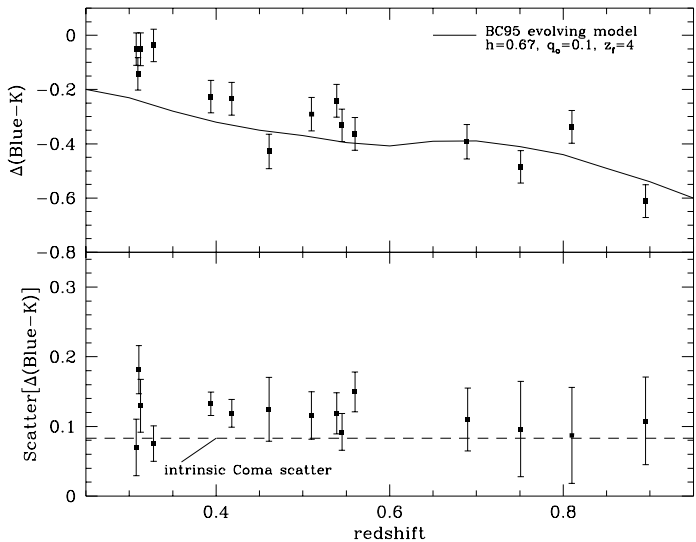,height=3in}}
\caption{Evolution of cluster ellipticals from $0.3 < z < 0.9$,
from Stanford, Eisenhardt and Dickinson 1997.  The galaxy sample
consists of morphologically selected E/S0 galaxies from 16 clusters
with 5--band optical--IR ground based and HST WFPC2 imaging.
The ``blue'' band shifts with cluster redshift so as always to
measure approximately rest--frame $U$--band light, while the 
$K$--band is fixed in the observed frame.  
{\it Top panel:} color evolution of the E/S0 population.  
The vertical axis represents the color {\it difference} relative 
to the same rest--frame wavelengths in the Coma cluster, i.e. 
``no evolution'' would be a horizontal line at $\Delta$(blue-$K$) = 0.
{\it Bottom panel:}  Scatter in E/S0 colors around the mean
color--magnitude locus.  The scatter in high--$z$ clusters is 
somewhat higher than for Coma, but changes negligibly with redshift
from $0.3 < z < 0.9$.}

\end{figure}

Although the evolutionary history of elliptical galaxies has been
the subject of debate, the evidence from rich clusters at $z < 1$ 
seems relatively unambiguous.  Bower, Lucey \& Ellis (1992) have 
cited the small color scatter in Coma and Virgo cluster ellipticals 
to argue for either a high redshift of formation, strongly synchronized 
coevality, or both.  At higher redshifts, early spectroscopic observations
(Dressler \& Gunn 1990), optical--IR colors (Arag\'on--Salamanca \etal 1993) 
and analysis of the fundamental plane and its projections 
(Van Dokkum \& Franx 1996;  Dickinson 1995; 
Pahre \etal 1996;  Schade \etal 1996) have all shown that the ``red envelope'' population 
of cluster ellipticals evolves slowly and in a fashion consistent 
with simple passive evolution and a high redshift of formation.
Recently, Stanford, Eisenhardt \& Dickinson (1997;  see also 
contribution by Stanford to this volume) have gathered deep, 
wide--field 5--band IR/optical 
imaging on 45 clusters from Coma out to $z=0.9.$  For 16 of these,
HST WFPC2 imaging allows us to morphologically select early--type 
galaxies from the cluster cores.  As seen in figure 1, the elliptical 
galaxy color--magnitude (c--m) relation shows only very mild color 
evolution out to $z=0.9$, and there is virtually no redshift evolution 
in the {\it scatter} of the galaxy colors around the mean c--m relation,
extending the Bower \etal result to large lookback times where the
{\it fractional} age differences between non--coeval galaxies would
be larger.  The c--m slope (not shown) also shows no significant change 
with redshift out to $z=0.9$.  These facts all point to a strongly coeval 
population and a high redshift of formation for cluster ellipticals, well 
beyond $z = 1$.   This is one strong motivation for our search for $z > 1$
galaxy clusters:  evidently by $z=0.9$ we have still not yet approached the 
epoch of star forming activity for most cluster ellipticals.

\begin{figure}
\centerline{\psfig{figure=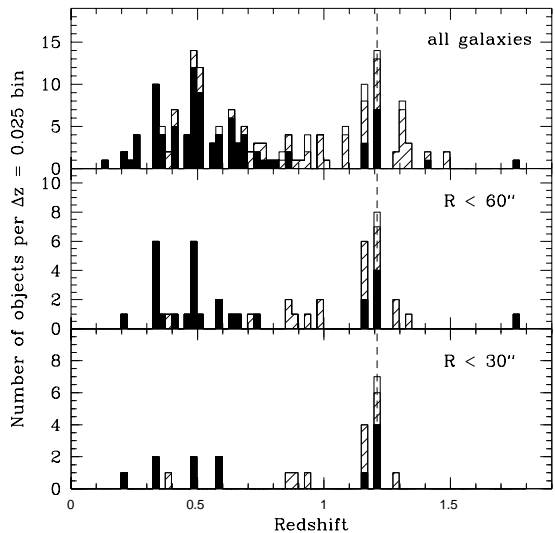,height=3in}}
\caption{Redshift distribution in the field of the $z=1.206$ radio
galaxy 3C~324.  The panels show subsamples restricted at several 
radii from the radio galaxy.  The shadings indicate redshift 
``quality classes;''  the filled area show redshifts from multiple 
spectral features, while the hatched redshifts are mostly based on
single emission lines, generally assumed to be [OII].  The vertical
dashed line marks the radio galaxy redshift.}
\end{figure}

Infrared imaging surveys offer the best promise for finding clusters
(particularly those rich in early--type galaxies) at $z > 1$.
Because of the current limitations imposed by the small size of 
infrared arrays, it is difficult to survey the large solid angles needed
to find rare objects like rich clusters.  We have chosen to narrow
the search to {\it a priori} likely sites, namely the environments of 
radio loud AGN.  Peter Eisenhardt and I have obtained deep $R$, $J$ and 
$K$ imaging of 25 powerful radio galaxies at $0.8 < z < 1.4$ and have 
identified several good candidates for rich clusters, primarily visible 
in the infrared data thanks to large overdensities of very red 
($R - K \approx 6$) galaxies with the colors expected for weakly evolved 
ellipticals at $z \simgt 1$.  Even here, however, spectroscopic 
confirmation is essential in order to sort out projection effects and 
study the galaxy population.

For one field from our sample, that around 3C~324 at $z=1.206$, 
Spinrad, Dey, Stern, LeF\`evre and I have measured $\sim 150$ galaxy 
redshifts down to $R=25$ and $K=20$ using Keck and the NTT.  3C~324 shows
one of the most prominent excesses of faint red galaxies in our infrared
imaging data.  The redshift distribution is shown in figure~2.  
Considering the entire $3^{\prime} \times 8^{\prime}$ survey field,
$>$80\% of the galaxies are foreground or background to the radio galaxy,
emphasizing the difficulty of identifying rich clusters at $z > 1$
even with extensive redshift data.  However, two prominent spikes are 
visible at $z = 1.15$ and 1.21, particularly when attention is restricted 
to radii within 1~arcmin of the radio galaxy.  Evidently, the 
``3C 324 cluster'' seen as an excess of galaxies in the near--IR divides 
into two distinct structures separated by $\sim 7500$~km~s$^{-1}$ in their 
rest frame.  The red ellipticals appear to be present in both structures.  

\begin{figure}[ht]
\centerline{\psfig{figure=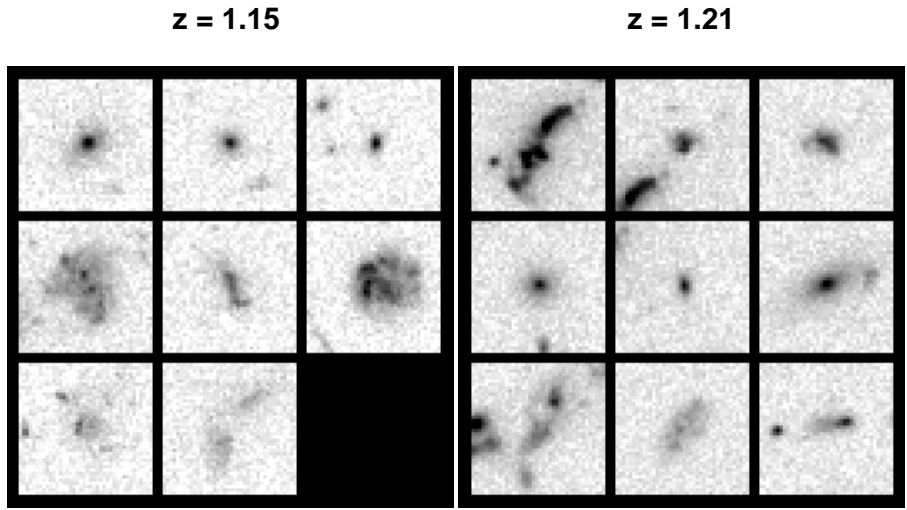,width=4.7in}}
\caption{WFPC2 F702W images of spectroscopically confirmed galaxies
in the 3C~324 cluster(s).  The galaxies at left are members of
the $z=1.15$ structure, including ellipticals, some irregular 
disks, and two large, diffuse galaxies (bottom).  Unusually, the 
bottom central object is extremely red and bright in the IR, 
suggesting perhaps a ``buried'' starburst.   The galaxies at right
are at $z=1.21$, including the radio galaxy (top left), several 
ellipticals, and peculiar objects (bottom).}
\end{figure}

With few exceptions, the red galaxies in the 3C~324 cluster(s) have 
simple elliptical morphologies.  Their $R-K$ colors are $\sim$0.6 magnitudes 
bluer than similar rest--frame colors for giant ellipticals today, 
consistent with an extrapolation of the color vs. $z$ relation for $z < 1$ 
cluster ellipticals found by Stanford {\it et al.}, and the scatter in their 
color--magnitude relation remains remarkably small, $\sim 0.07$ magnitudes.   
The rest--frame ultraviolet spectra of the 3C~324 ellipticals are 
qualitatively very similar to that of the red $z \approx 1.5$ radio galaxies
53W091 and 53W069 (see Dunlop \etal 1996, and Dey, this volume).  
Their surface brightnesses
suggest $\sim$1 magnitude of rest--frame $B$--band luminosity evolution
(cf. Dickinson 1995), fully consistent with passive evolutionary models.
The other spectroscopically confirmed cluster members (figure~3) exhibit
a bewildering range of morphologies.  Very few, if any, ``normal'' disk
galaxies have been identified, although the degree to which this is a 
consequence of the ultraviolet rest--frame wavelengths imaged by 
WFPC2 at $z=1.2$ (3200\AA\ for the F702W filter) is unclear:  NICMOS 
imaging planned as a HST Cycle~7 GTO program should help clarify this.

\begin{figure}[ht]
\centerline{\psfig{figure=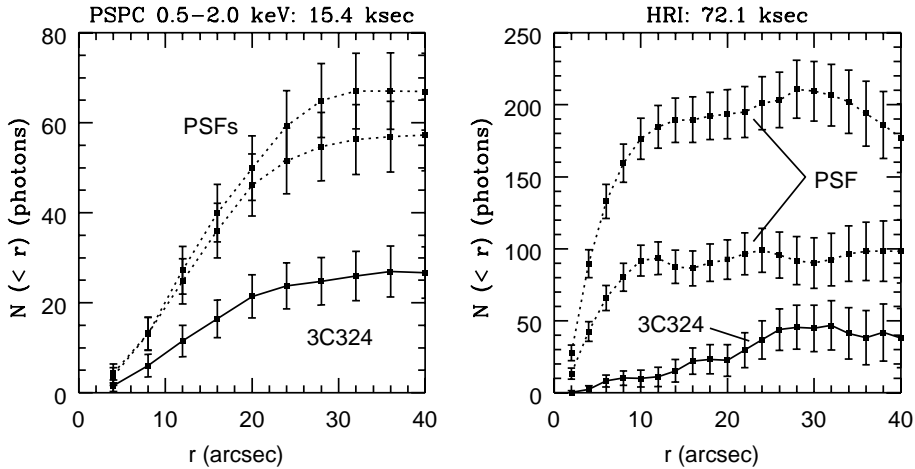,height=2.7in}}
\caption{Radial curves of growth for x--ray emission around
3C~324 from the Rosat PSPC (left) and HRI (right).  The dashed curves
labeled PSF show the profiles of two point sources nearby in the field.
In the low angular resolution PSPC data, 3C~324 is not convincingly
resolved, but the HRI data shows that the x--rays are extended over 
a detectable diameter of $\sim 60^{\prime\prime}$, strongly suggesting 
the presence of a luminous x--ray cluster associated with the radio galaxy.}
\end{figure}

We have also found evidence that the 3C~324 cluster is 
a {\it massive, collapsed system.}  
In the course of a Rosat PSPC survey of $z > 1$ radio galaxies, we 
detected a faint x--ray source coincident (within $10^{\prime\prime}$)
of 3C~324.  The poor angular resolution of the PSPC did not allow us
to resolve the x--ray emission at the low signal--to--noise
of the source, however, leaving ambiguous whether the x--rays arose from
the radio galaxy AGN or from surrounding cluster gas.  A subsequent
72.1 ksec Rosat HRI exposure, however, clearly shows that the x--ray emission
is resolved over a detectable diameter of $\sim 60^{\prime\prime}$
($\sim 0.5$ Mpc) (figure 4).   Our spectroscopy shows no other
substantial galaxy groups along the line of sight so closely aligned with
the radio/x--ray source.  This strongly suggests that the x--ray emission 
comes from resolved cluster gas at $z \approx 1.2$ with a bolometric 
$L_X = (8.1\pm1.6) \times 10^{44}$~erg~s$^{-1}$ (for $H_0 = 50$,
$q_0 = 0.5$), comparable to that of the Coma cluster.
We have subsequently resolved two additional radio galaxy x--ray 
sources at $z > 1$ with the Rosat HRI, and Crawford \& Fabian (1996)
have resolved a fourth.  It seems quite unlikely that these multiple 
examples of extended x--ray emission around radio galaxies could all
arise from foreground group/cluster sources seen in projection.  
The most distant of 
our resolved HRI detections is for 
a radio galaxy at $z=1.8$ with $L_X \approx 5 \times 10^{44}$~erg~s$^{-1}$.
Unless other explanations for highly extended x--ray emission with 
these luminosities can be concocted, it seems clear that some radio galaxies
out to nearly $z=2$ were situated in deep gravitational potential wells
comparable to those of massive clusters today, providing a challenge
for theories of cosmic structure formation.

\section{Clusters at $z > 2$}

Although the evidence for clusters at $z > 2$ is extremely heterogeneous,
a number of observations have been presented recently
which suggest that the idea is not pure fantasy.  Most importantly, 
recent spectroscopic confirmation of galaxy redshifts in several examples
(cf. Francis \etal 1996, Pascarelle \etal 1996, Malkan \etal 1996, 
M\o ller \& Warren 1993, LeF\`evre et al. 1996) has lent credence to 
earlier claims based on statistical 
overdensities or narrow band imaging studies alone.   Table~1 below
presents an almost certainly incomplete summary of the more interesting
examples of suggested (and in many instances confirmed) 
galaxy associations at $z > 2$.  

Most impressively, Pascarelle \etal (1996) report spectroscopic redshifts
for 5 objects out of 18 candidates in an apparent cluster at $z=2.39$,
selected by Ly$\alpha$ excess in intermediate bandwidth HST images.
Windhorst (priv. comm.) reports that at least four further members have
been confirmed with recent Keck spectroscopy.  The cluster objects
are very compact, like most very faint galaxies seen in deep WFPC2 images.
Curiously, many of them show AGN--like spectral features (CIV, NV emission,
etc.).   Objects in several other $z > 2$ cluster candidates also show 
AGN features or radio emission (Malkan \etal 1996;  LeF\`evre \etal 1996; 
Francis \etal 1996) suggesting that AGN activity may have been quite
prevalent in these environments and/or at these redshifts.  In part, this 
may be a consequence of galaxy selection from narrow band emission line
imaging, especially with Lyman~$\alpha$:  evidence from high--$z$ field 
galaxies indicates that Ly$\alpha$ emission is typically very weak or 
absent in star--forming galaxies, presumably due to effects of dust and 
resonant scattering.

At present, the relation of these $z > 2$ associations to nearby
rich clusters is unknown.   In particular, it is unclear whether these
are massive, collapsed systems, groupings within unvirialized ``sheets''
of galaxies, or collections of ``protogalactic'' fragments
destined to merge into single, more massive galaxies.   But the
range of new data is impressive and suggestive, and further investigations
should be fruitful.  With the dramatic progress of Keck spectroscopy 
on extremely faint high--$z$ galaxies, and with the promise of 
infrared OH suppression spectrographs in the near future which will
facilitate redshift measurement for galaxies at $z > 1.2$, 
we may fully expect this evidence to continue to accumulate.

\begin{table}[htp]
\caption{Cluster candidates at $z > 2$}
\vspace{0.4cm}
\begin{center}
\begin{tabular}{|c|l|l|}
\hline 
$z$ & Reference & Description\\
\hline 
2.05 & Dressler \etal 1993 &
\begin{minipage}[t]{2.3in}
QSO with local overdensity of peculiar objects in HST images 
\end{minipage}\\&&\\
2.38 & Francis \etal 1996 &
\begin{minipage}[t]{2.3in}
Multiple QSO absorbers at common $z$ in separate sightlines;
Ly$\alpha$ excess objects; some spectroscopic confirmation;
red IR--IR color selected objects 
\end{minipage}\\&&\\
2.39 & Pascarelle \etal 1996 &
\begin{minipage}[t]{2.3in}
$\sim$18 Ly$\alpha$ selected compact galaxies in deep WFPC2 
images; $\sim$9 confirmed spectroscopically 
\end{minipage}\\&&\\
$\sim 2.5$ & \begin{minipage}[t]{1.5in} 
Arag\'on--Salamanca {\it et~al.~} 1996 
\end{minipage}&
\begin{minipage}[t]{2.3in}
Statistical $K$--band excess around radio loud quasars 
\end{minipage}\\&&\\
2.50 & Malkan \etal 1996 &
\begin{minipage}[t]{2.3in}
3 H$\alpha$ selected (in infrared) galaxies at redshift of
known damped Ly$\alpha$ QSO absorber
\end{minipage}\\&&\\
2.81 & \begin{minipage}[t]{1.5in}
M\o ller \& Warren 1993, 
Warren \& M\o ller 1996 
\end{minipage}&
\begin{minipage}[t]{2.3in}
3 spectroscopically confirmed, Ly$\alpha$ selected compact 
galaxies (HST imaging) associated with $z_{abs} = z_{QSO}$
damped Ly$\alpha$ absorber and QSO. 
\end{minipage}\\&&\\
3.14 & LeF\`evre \etal 1996 &
\begin{minipage}[t]{2.3in}
2 spectroscopically confirmed Ly$\alpha$ selected companions
to a powerful radio galaxy 
\end{minipage}\\&&\\
3.58 & Giavalisco \etal (in prep.) &
\begin{minipage}[t]{2.3in}
Excess of Lyman break selected galaxies in field of powerful
radio galaxy 
\end{minipage}\\&&\\
3.80 & Lacy \& Rawlings 1996 &
\begin{minipage}[t]{2.3in}
Excess of Lyman break selected galaxies in field of powerful
radio galaxy 
\end{minipage}\\&&\\
4.55 & Hu \& McMahon 1996 &
\begin{minipage}[t]{2.3in}
2 Ly$\alpha$ selected objects associated with QSO 
\end{minipage}\\&&\\
\hline 
\end{tabular}
\end{center}
\end{table}

\section*{Acknowledgements}

I would like to extend special thanks to my collaborators
(particularly Adam Stanford, Peter Eisenhardt, Hy Spinrad, Arjun Dey, 
Daniel Stern, Olivier LeF\`evre, and Richard Mushotzky) for permitting
me to present data prior to publication, and to the conference organizers
for their invitation and travel support.

\end{document}